\documentclass[fleqn,usenatbib]{mnras}

\usepackage{mathptmx}

\usepackage[T1]{fontenc}

\DeclareRobustCommand{\VAN}[3]{#2}
\let\VANthebibliography\thebibliography
\def\thebibliography{\DeclareRobustCommand{\VAN}[3]{##3}\VANthebibliography}

\usepackage{graphicx}   
\usepackage{amsmath}    
\usepackage{amssymb}    

\title[Role of the radiative stage for cosmic ray acceleration in SNRs]{Role of the radiative stage for cosmic ray acceleration in SNRs}

\author[V. N. Zirakashvili and V.S.Ptuskin]{
V. N. Zirakashvili,\thanks{E-mail: zirak@izmiran.ru} and V. S.
Ptuskin,
\\
Pushkov Institute of Terrestrial Magnetism, Ionosphere and Radiowave
Propagation, 108840, Troitsk, Moscow, Russia}

\date{Accepted 2021 December 15. Received 2021 December 12; in original form 2021 November 20}

\pubyear{2021}

\begin{document}
\label{firstpage}
\pagerange{\pageref{firstpage}--\pageref{lastpage}} \maketitle

\begin{abstract}
We consider diffusive shock acceleration in supernova remnants
throughout their evolution including a radiative stage. It is found
that a more efficient acceleration and fast exit of particles at the
radiative stage results in the hardening
 of the source cosmic ray proton and electron spectra at energies $\sim 100-500$ GeV.
The effect is stronger for cosmic ray electrons.
\end{abstract}




\section{Introduction}

Supernova remnants (SNRs) are considered now as a principle source
of Galactic cosmic rays (CRs). It is believed that the diffusive
shock acceleration  (DSA) mechanism
 \citep{krymsky77,bell78,axford77,blandford78} operates
 in the vicinity of shocks in SNRs. During the last decades, the modern
X-ray and gamma-ray observations supplied the evidence of the
presence of multi-TeV energetic particles in these astrophysical
objects (see e.g. \citet{lemoine14} for a review).

Usually the existing DSA models are applied for young SNRs where the
most energetic CRs  are accelerated. However, lower energy particles
are produced  in older SNRs either. The investigation of CR
acceleration in SNRs throughout all evolutionary stages is important
for the calculation of overall CR spectra produced
 by SNRs.

In this paper, we describe the modifications of our non-linear
 DSA model \citep{zirakashvili12} designed for investigation of DSA over the entire life of SNRs. The preliminary results
 on the application of this modified model to gamma ray bright SNRs W28, W44, and IC443 were reported in
\citet{zirakashvili18,zirakashvili16a,zirakashvili18a}. The main new
features are the gas ionization by the radiation of the remnant,
radiative cooling of the gas, and damping of MHD waves on neutral
atoms at the late stages of SNR evolution. We also performed the
modeling of acceleration and production of broad-band
electromagnetic emission in the young SNR Tycho and middle-aged SNR
W44 to adjust the parameters of the model.

The paper is organized as follows. In the next Section 2, we
describe our model. CR acceleration in IIP type SNRs evolving in the
dense medium and modeling of nonthermal emission of SNR W44 is
presented in Section 3. The modeling of Ia Type SNRs evolving in a
more rarefied medium
 and the modeling of SNR Tycho are described in Section 4.
The discussion of results and conclusions are given in Sections 5
and 6.

\section{Nonlinear diffusive shock acceleration model}

Details of our basic model of nonlinear DSA can be found in
\citet{zirakashvili12}. The model contains coupled spherically
symmetric hydrodynamic equations  and the transport equations for
energetic protons, ions, and electrons. The forward and reverse
shocks are included in the consideration.

Damping of magnetohydrodynamic (MHD) waves due to the presence of
neutral atoms is important for SNRs expanding in not fully ionized
gas. To take
 this effect into account we add the equation that describes the transport and generation of MHD waves
(see Eq.(5) below).

The hydrodynamical equations for the gas density  $\rho (r,t)$, gas
velocity $u(r,t)$, gas pressure $P_g(r,t)$, wave pressure
$P_w(r,t)$, pressure of the regular magnetic field $P_m(r,t)$, and
the equation for isotropic part of the cosmic ray proton momentum
distribution
 $N(r,t,p)$ in the spherically symmetrical  case are given by

\begin{equation}
\frac {\partial \rho }{\partial t}=-\frac {1}{r^2}\frac {\partial
}{\partial r}r^2u\rho
\end{equation}

\begin{equation}
\frac {\partial u}{\partial t}=-u\frac {\partial u}{\partial
r}-\frac {1}{\rho } \left( \frac {\partial P_g}{\partial r}+\frac
{\partial P_c}{\partial r} +\frac {\partial P_w}{\partial r}+\frac
{\partial P_m}{\partial r}\right)
\end{equation}

\[
\frac 1{\gamma _g-1}\left( \frac {\partial P_g}{\partial t}+u\frac
{\partial P_g}{\partial r} +\frac {\gamma _gP_g}{r^2}\frac {\partial
r^2u}{\partial r}\right) =
\]
\begin{equation}
-\Lambda (T_e)n^2+H_{c}+2\Gamma _n\frac {P_w}{\gamma _w-1}-\xi
_AV_{Ar}(1-h_w)\frac {\partial P_c}{\partial r}
\end{equation}

\begin{equation}
\frac {\partial P_m}{\partial t}+u\frac {\partial P_m}{\partial r}
+\frac {\gamma _mP_m}{r^2}\frac {\partial r^2u}{\partial r}=0
\end{equation}

\[
\frac {\partial P_w}{\partial t}+(u+\xi _AV_{Ar})\frac {\partial
P_w}{\partial r} +\frac {P_w}{r^2}\frac {\partial r^2(\gamma _wu+\xi
_AV_{Ar})}{\partial r}=
\]
\begin{equation}
-h_w(\gamma _w-1)\xi _AV_{Ar}\frac {\partial P_c}{\partial
r}-2\Gamma _nP_w
\end{equation}

\[
\frac {\partial N}{\partial t}=\frac {1}{r^2}\frac {\partial
}{\partial r}r^2D(p,r,t) \frac {\partial N}{\partial r} -w\frac
{\partial N}{\partial r}+\frac {\partial N}{\partial p} \frac
{p}{3r^2}\frac {\partial r^2w}{\partial r}
\]
\[
+\frac 1{p^2}\frac {\partial }{\partial p}p^2b(p)N+
\]
\[
\frac {\eta _fX_i\delta (p-p_{f})}{4\pi p^2_{f}m}\rho
(R_f+0,t)(\dot{R}_f-u(R_f+0,t))\delta (r-R_f(t))
\]
\begin{equation}
+\frac {\eta _b\delta (p-p_{b})}{4\pi p^2_{b}m}\rho
(R_b-0,t)(u(R_b-0,t)-\dot{R}_b)\delta (r-R_b(t))
\end{equation}
Here $P_c=4\pi \int dpp^3vN/3$ is the cosmic ray pressure, $w(r,t)$
is the advection velocity of cosmic rays, $T_e$, $\gamma _g$ and $n$
are the gas temperature, adiabatic index, and number density
respectively,  $\gamma _w$ is the wave adiabatic index, $D(r,t,p)$
is the cosmic ray diffusion coefficient. The radiative cooling of
gas is described by the cooling function $\Lambda (T_e)$.
 The function $b(p)$ describes the energy losses of particles.
In particular, the Coulomb  losses of sub-GeV ions
  and the radiative cooling are important in old SNRs. The energy of
 sub GeV ions goes to the gas heating described by the term $H_c$ in Eq. (3).

Cosmic ray diffusion is determined by the scattering on magnetic
inhomogeneities. The cosmic ray streaming instability increases the
level of MHD turbulence in the shock vicinity \citep{bell78} and
even significantly amplifies the absolute value of the magnetic
field in young SNRs \citep{bell04,zirakashvili08}. It decreases the
diffusion coefficient and increases the maximum energy of
accelerated particles. The results of continuing theoretical study
of this effect can be found in review papers
\citep{bell2014,Caprioli2014}.

Cosmic ray particles are scattered by moving waves and it is why the
cosmic ray advection velocity
 $w$ may differ from the gas velocity $u$ by the value of the radial
component of the Alfv\'en velocity $V_{Ar}=V_A/\sqrt{3}$ calculated
in the isotropic random magnetic field: $w=u+\xi _AV_{Ar}$. The
factor $\xi _A$ describes the possible deviation of the cosmic ray
drift velocity from the gas velocity. We  use values $\xi _A=1$ and
$\xi _A=-1$ upstream of the forward and reverse shocks respectively,
where Alfv\'en waves are generated by the cosmic ray streaming
instability and propagate in the corresponding directions.

The situation is less clear in the downstream region of the shocks.
Usially, the Alfv\'en drift
 is not considered here. However, it is known that Alfv\'en transport in the
downstream region suggested at phenomenological level
\citep{zirakashvili08b} results in steeper spectra of accelerated
particles. This also allow avoiding a cosmic ray overproduction in
evolutionary models of SNRs \citep{ptuskin10}.

Recently the Alfv\'en transport in the downstream region was indeed
observed in hybrid modeling of collisionless shocks
\citep{haggerty20}. It looks like some nonlinear magnetic structures
are generated in  the shock transition and move with Alfv\'en speed
in the downstream region. The origin of these
 nonlinear waves is unclear. They can be large scale transverse Alfv\'en-like waves propagating
in the isotropic tangle magnetic field with phase speed
$V_A/\sqrt{3}$ \citep{moffatt86}. Or they somehow can be related to
sonic waves generated at the shock front.

Below we take the Alfv\'enic transport in the downstream region into
account. We use values $\xi _A=-1$ and $\xi _A=1$ just downstream of
forward and reverse shocks respectively. The Alv\'enic transport
takes place in the narrow region of thickness 0.1 of the distance
between the shock and the contact discontinuity.

The pressure of generated waves $P_w$ determines the scattering and
diffusion of
 energetic particles with charge $q$, momentum $p$, and speed $v$

\begin{equation}
D=D_B\frac {P_m+P_w}{P_w}, \ D_B=\frac {cpv}{3qB}, \ B=\sqrt{8\pi
(P_m+P_w)}
\end{equation}
where $B$ is the total magnetic field strength, while $P_m$ is the
pressure of the regular field. At high wave amplitudes, the
diffusion coefficient coincides with the  Bohm diffusion coefficient
$D_B$.

The parameter $h_w$ in Eqs. (3,5)  describes the fraction of the
wave energy
  produced by the streaming instability. We use the following dependence $h_w(P_w/P_m)$

\begin{equation}
h_w=1, \ \frac {P_w}{P_m}<10; \ h_w=0.7, \ \frac {P_w}{P_m}>10.
\end{equation}
At high amplitudes, the waves are damped and the fraction $1-h_w$ of
energy goes into
 the gas heating
upstream of the shocks \citep{mckenzie82} that is described by the
last term in Eq. (3). The heating and wave generation limits the
total compression ratio of cosmic ray modified shocks. The value of
$h_w$ regulates the magnetic amplification in the upstream region of
the shock. Since
 the amplified field is transported into the downstream region, $h_w$ also determines the efficiency of
 the Alfv\'enic transport in this region and regulates the spectral slope of accelerated particles.
Its value $h_w=0.7$ was adjusted to reproduce broad-band modeling of
Tycho SNR (see Section 4 below).

The low seed level of the interstellar turbulence
$P_w=10^{-6}B_0^2/8\pi $ is prescribed at the simulation
 boundary at $r=2R_f$. Since the flux of escaped highest energy particles  amplifies
waves exponentially in time the results depend only logarithmically
on the seed level.

 In the shock transition region the wave pressure
 is increased by a  factor of $\sigma ^{\gamma _w}$,  where
$\sigma $ is the shock compression ratio. Its impact  on the shock
dynamics is taken into account via the Hugoniot conditions.

Below we use the adiabatic index of Alfv\'en waves $\gamma _w=3/2$.
For this value of the adiabatic index, the wave pressure
$P_w=(\delta B)^2/8\pi $ equals
 the  wave magnetic energy density.
The pressure of the regular field $P_m$ plays a dynamical role
at the radiative phase when the field is strongly compressed in the
downstream region and produces a significant anisotropic force in
the radial direction. To take this into account we use the
 adiabatic index $\gamma _m=2$ for the regular magnetic field.

The rate of the neutral damping $\Gamma _n=0.5\nu _{in}$ is
determined by the frequency of ion neutral collisions $\nu _{in}$ in
the limit of the high wave frequencies $\omega >> \nu _{in}$. The
frequency of ion-neutral collisions is determined by charge-exchange
process and is  given by (\citet{drury96})
\begin{equation}
\nu _{in}=8.9\cdot 10^{-9} n_n \left( \frac {T}{10^4\
\mathrm{K}}\right) ^{0.4}\ \mathrm{s}^{-1}.
\end{equation}
Here the number density $n_n=X_nn_H$ of neutral hydrogen atoms is
determined by the neutral fraction $X_n$. A useful review of DSA in
partially ionized plasma can be found in \citet{bykov13}.

The neutral fraction of hydrogen ions $X_n$ is determined by
equation
\begin{equation}
\frac {\partial X_n}{\partial t}=-u\frac {\partial X_n}{\partial r}
+\alpha _{\mathrm{rec}}n_HX^2_i-X_n(q_{\mathrm{th}}+q_{\mathrm{ph}})
\end{equation}
where $X_i=1-X_n$ is the fraction of ionized Hydrogen, $\alpha
_{\mathrm{rec}}$ is the recombination rate and $q_{\mathrm{th}}$ and
$q_{\mathrm{ph}}$ are thermal ionization and photoionization rates
respectively. The photoionization rate
 is given by

\begin{equation}
q_{\mathrm{ph}}(r)= 2\pi \int ^\pi _0\sin (\theta )d\theta \sigma
_{\mathrm{ph}}I(r,\theta )
\end{equation}
Here $\theta $ is the angle between the photon wavevector and radial
direction and $\sigma _{\mathrm{ph}}$ is the photoionization
cross-section of Hydrogen. The intensity of ionizing photons
$I(r,\theta )$ is determined from the equation of
 radiative transport
\begin{equation}
\cos (\theta )\frac {\partial I}{\partial r} -\frac {\sin (\theta
)}r\frac {\partial I}{\partial \theta} = \frac {\Lambda _i(T_e)
n^2}{4\pi I_H}-\sigma _{\mathrm{ph}}I
\end{equation}
and $I_H=13.6$ eV is the ionization potential of Hydrogen. The first
term  on the right-hand side is the emissivity of ionizing photons
that was determined by the partial cooling function $\Lambda _i(T)$
in the radiation range
 300-910 angstrom of \citet{landini90}.

Two last terms in Eq. (6) correspond to the injection of thermal
protons with momenta $p=p_{f}$, $p=p_{b}$ and mass $m$ at the
forward and reverse shocks located at $r=R_f(t)$ and $r=R_b(t)$
respectively. The dimensionless parameters $\eta _f$ and $\eta _b$
determine the efficiency of injection.

The injection efficiency is taken to be independent of time $\eta
_f=0.001$, and the particle injection momentum is
$p_{f}=2m(\dot{R}_f-u(R_f+0,t))$. Protons of mass $m$ are injected
at the forward shock and ions of mass $M$ and mass to charge ratio
$A/Z=2$ and
 injection efficiency $\eta _b=0.001$ are injected at the reverse shock.

For high Mach number shocks this injection efficiency and Alfv\'en
transport in the downstream region limit the pressure of accelerated
particles and the magnetic energy density at the level
  $10-20\ \%$ and $1.5\ \%$ of the ram pressure of the shock respectively.
These numbers are comparable with ones observable in the hybrid
modeling of collisionless shocks \citep{caprioli20}.

We neglect the pressure of energetic electrons and treat them as
test particles. The evolution of the electron distribution is
described by the equation analogous to Eq. (6) with function $b(p)$
describing Coulomb, synchrotron, and inverse Compton (IC) losses and
additional terms describing the production of secondary leptons by
energetic protons and nuclei. The electron injection efficiency
$\eta _e$ at the forward shock was taken in the form
\begin{equation}
\eta _e=7\cdot 10^{-8}\left( \frac {V_f}{c}\right) ^{-0.6}
\end{equation}
where the numeric parameters were adjusted to reproduce the
intensity of radio-emission in supernova remnants W44 and Tycho.
This dependence of $\eta _e$ on the shock velocity $V_f$ results in
a higher electron to proton ratio in older SNRs in comparison with
the one in the young SNRs.


\section{Modeling of diffusive shock acceleration in SNR of IIP supernova}

A significant part of core-collapse supernova explosion occurs in
molecular gas. The stars with
 initial masses below $12\ M_{\odot}$ have no power stellar winds and therefore do not produce a strong
 modification of their circumstellar medium. The molecular cloud has been totally destroyed by stellar winds and supernova
 explosions  of more massive stars at the instant of explosion. As a result, the star explodes in the inter-clump
 medium with the density $5-25$ cm$^{-3}$ \citep{chevalier99}.  Many such SNRs are observed in gamma rays now.

\begin{table*}[t]
\begin{center}
\caption{Physical parameters of SNRs modeled}
\begin{tabular}{|c|c|c|c|c|c|c|c|c|c|c|c|c|c|c|}
\hline  &Type&SNR  &$d$  & $R_f $& $E_{SN}$     & $M_{ej}$   & $k_{SN}$  &$B_0$& $n_H   $&$X_{i0}$&$R_i$& $T$&$V_f$& $B_f$\\
\hline  &    &     &kpc  & pc    & $10^{51}$erg &$M_{\odot }$&        &$\mu $G &cm$^{-3}$&        & pc  &kyr &km/s &$\mu $G\\
\hline  &IIP &W44  &2.8  & 12.35  & 1.8          & 10         &  9     &5      & 7.0     & 0.01   &  4  & 23 &160  &  51\\
\hline  &Ia  &Tycho&3.5  & 4.1    & 1.2          & 1.4        &  7     &3      & 0.2     & 0.5    &  8  &0.44&5300 & 210   \\
\hline
\end{tabular}
\end{center}
\end{table*}

\begin{figure}
\begin{center}
\includegraphics[width=8.0cm]{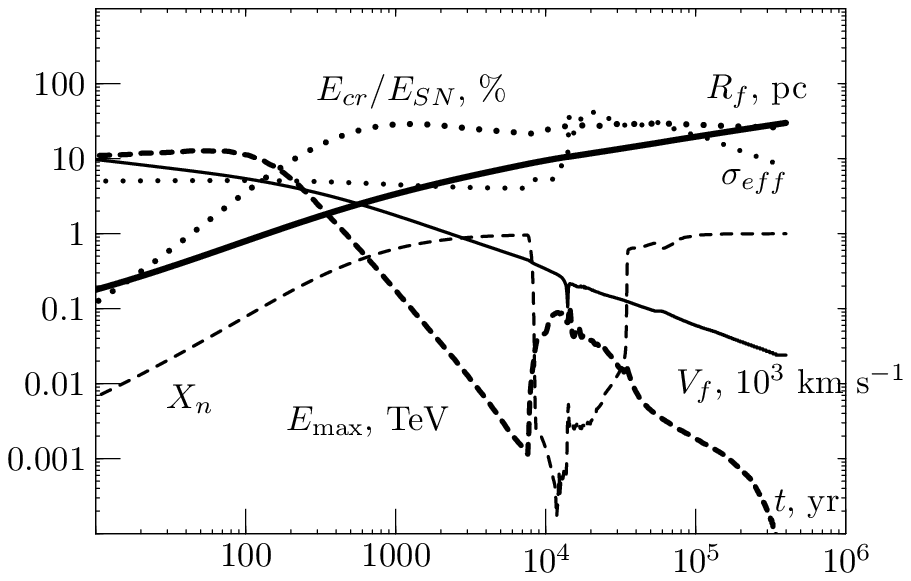}
\end{center}
\caption{ Dependence on time of the forward shock radius $R_f$
(thick solid line), the shock speed $V_f$
 (thin solid line), the shock effective compression ratio $\sigma _{eff}$ (thin dotted line),
the maximum energy of particles accelerated at forward shock
$E_{\max }$ (dashed line), the fraction of explosion energy
transformed into cosmic rays $E_{cr}/E_{SN}$ (thick dotted line) and
neutral fraction $X_n$ (thin dashed line) calculated for SNR of IIP
supernovae.}
\end{figure}

It is believed that the circumstellar medium  is almost fully
 ionized by ultraviolet radiation from the remnant interior at the radiative stage \citep{chevalier99}.
The same is true for young SNRs
 because the gas is ionized by the radiation from the shock breakout and in the hot shock precursor produced
by accelerated particles. Extended red super giant progenitors of
IIP Type supernovae emit  $\sim 10^{48}$ erg of radiation during the
shock breakout. This amount of energy is sufficient for the
ionization of several dozens of solar masses of
 the circumstellar gas \citep{chevalier05}. If so the shock propagates in the preionized medium at the free
expansion phase and at the beginning of
 the Sedov stage. The amount of the breakout radiation is significantly smaller for compact progenitors
of Ib/c and Ia type supernovae. However Wolf-Rayet progenitors
itself ionize $\sim 10^3M_{\odot }$ of surrounding gas before the
explosion. Some level of the preionization is also expected for Ia
Type supernovae because of the ionizing radiation of the accreting
white dwarf. In this regard the assumption of the full ionization is
justified for almost  all stages of the supernova remnant evolution.
The only probable exception is the end of the Sedov stage when the
shock might propagate in the neutral medium. In this picture, the
number density of neutrals $n_n$ is determined by the ionization
history and by the recombination in ionized or preionized plasma.
\begin{figure}
\begin{center}
\includegraphics[width=8.0cm]{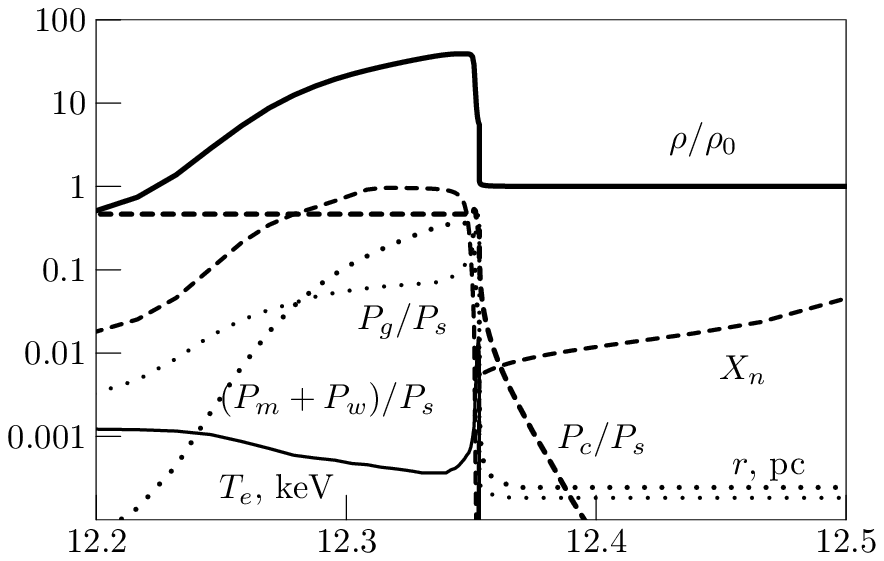}
\end{center}
\caption{Radial dependencies of the gas density (thick solid line),
the gas temperature $T_e$ (thin solid line), CR pressure (thick
dashed line), neutral fraction $X_n$ (dashed line), the magnetic
energy density $B^2/8\pi =P_m+P_w$ (dotted line), and the gas
pressure $P_g$ (thin dotted line) at
 $T=23$ kyr in the vicinity of the forward shock in SNR W44. The pressures are normalized to the ram pressure
of the shock $P_s=\rho _0V^2_f$}
\end{figure}

Bright in GeV gamma-rays middle-aged SNR W44 at distance $d=2.8$ kpc
from the Earth is at the radiative phase now and shows signs of
interaction with molecular gas \citep{reach05}. It contains HI shell
expanding with speed $135-150$ km s$^{-1}$ \citep{koo95, park13}.

The parameters of our supernova modeling are given in Table I. The
explosion parameters were adjusted to reproduce
 broad-band observations of  SNR W44.
The explosion energy $E_{SN}$ and ambient number density $n_H$ were
adjusted to reproduce the expansion speed of HI shell and the
observable gamma-ray flux.

The numbers in the three last columns of Table I that is the age
$T$, shock speed $V_f$ and magnetic field strength $B_f$  just
downstream of the shock were obtained in the modeling.

We use the parameter of ejecta velocity distribution $k_{SN}=9$
(this parameter describes the power-law density profile
$\rho_{ej}\propto r^{-k_{SN}}$ of the outer part of the ejecta that
freely expands after supernova explosion).

The ionized fraction of Hydrogen $X_i$ at the initial instant of
time  was taken in the form
\begin{equation}
X_i=(1-X_{i0})\exp \left( -\left( \frac r{R_i}\right) ^2 \right)
+X_{i0}
\end{equation}
where the ionization fraction at infinity $X_{i0}$ and the radius of
the ionization zone $R_i$ are given in Table I. It was assumed that
about 40 solar masses of Hydrogen is ionized during the supernova
explosion.


Figures (1)-(4) illustrate the results of our numerical
calculations.

The temporal evolution of the remnant and particle acceleration are
illustrated in Fig.1. The maximum energy of particles
 $E_{\max }$ at the forward shock was estimated as the energy where the function $p^5N(p)$ has a maximal value.
The shock speed is almost constant  at the initial free expansion
stage. After one hundred years the Sedov stage begins. Maximum
energy of particles is close to 20 TeV at this time.
 Later the maximum energy drops sharply because of the shock velocity decrease  and because
of the neutral damping of MHD waves.
 The neutral fraction $X_n$ increases because of the recombination and later when the shock enters
into a neutral medium. Strictly speaking the adequate description of
DSA at this stage requires a kinetic treatment for the transport of
neutral atoms near the shock (see \citet{morlino13}). We leave the
 detailed description of the acceleration at this stage to the future. It seems that this stage does not produce
 a strong impact on results because the injection rate  is proportional to the ionized fraction $X_i$ and
 therefore the production rate of cosmic rays at this phase is not high.

The ionizing radiation from the shock interior again ionizes the
medium after $8$ kyrs. At this instant of time, the boundary of the
ionization zone overtakes the forward shock. The acceleration
efficiency encreases because of higher injection rate.

Several thousands years after this the  radiative stage begins  when
the cooling behind the shock results in the gas  compression in this
region and in the formation of a dense shell at the age T$=14$ kyrs.
While the compression ratio of the shock is close to the standard
value $\sim 4$
 the gas density continues to increase further downstream. To illustrate this we show in Fig.1 the
effective compression ratio $\sigma _\mathrm{eff}$ that is the ratio
of the maximum gas density behind the shock to the ambient gas
density. This ratio increases up to the value of 40 that is limited
by the presence of
 cosmic rays and regular magnetic fields.
This gas compression results in a more
 efficient acceleration and in an enhanced flux of runaway highest energy particles.
This, in turn,  produces some temporary constant level of the
maximum energy. In addition, the gas cooling behind the shock is
 accompanied by its recombination. The  neutral gas of the shell absorbs
the ionizing radiation from the hot remnant interior.  This stops
ionization in the far upstream region and gas ionized
 earlier begins to recombine. Particles accelerated earlier leave the neutral shell because of the damping of MHD waves.
At 35 kyrs the shock reaches the boundary of the ionized region and
enters the neutral medium. At later times the acceleration at the
forward shock does not occur.

It should be noted that we use a simplified approach for the
description of the radiative cooling and photoionization with
 the equilibrium cooling functions $\Lambda(T_e)$ and $\Lambda _i(T_e)$. In reality, radiative cooling and
photoionization depend on the ionization state of ions in plasma
that is not in thermal equilibrium. However, we
 checked by performing a test run without cosmic rays that the reionization, the dense shell formation,
and exit of the shock to the neutral
 medium occur at the same age as in the SNR modeling with a full application
of the atomic physics \citep{sarkar21}.

Radial dependencies of physical quantities in SNR W44 at present
 ($T=23$ kyr) are shown in Fig.2. The gas temperature drops
sharply downstream of the forward shock due to the radiative cooling
and a thin neutral dense shell is formed behind the forward shock.
We obtain the shell mass of the neutral Hydrogen 640 $M_\odot $ that
is somewhat higher than
 the measured value of 390 $M_\odot $\citep{park13}.
The central part of the remnant is filled by the hot rarefied gas
with a temperature of $10^6-10^7$K.

\begin{figure}
\includegraphics[width=8.0cm]{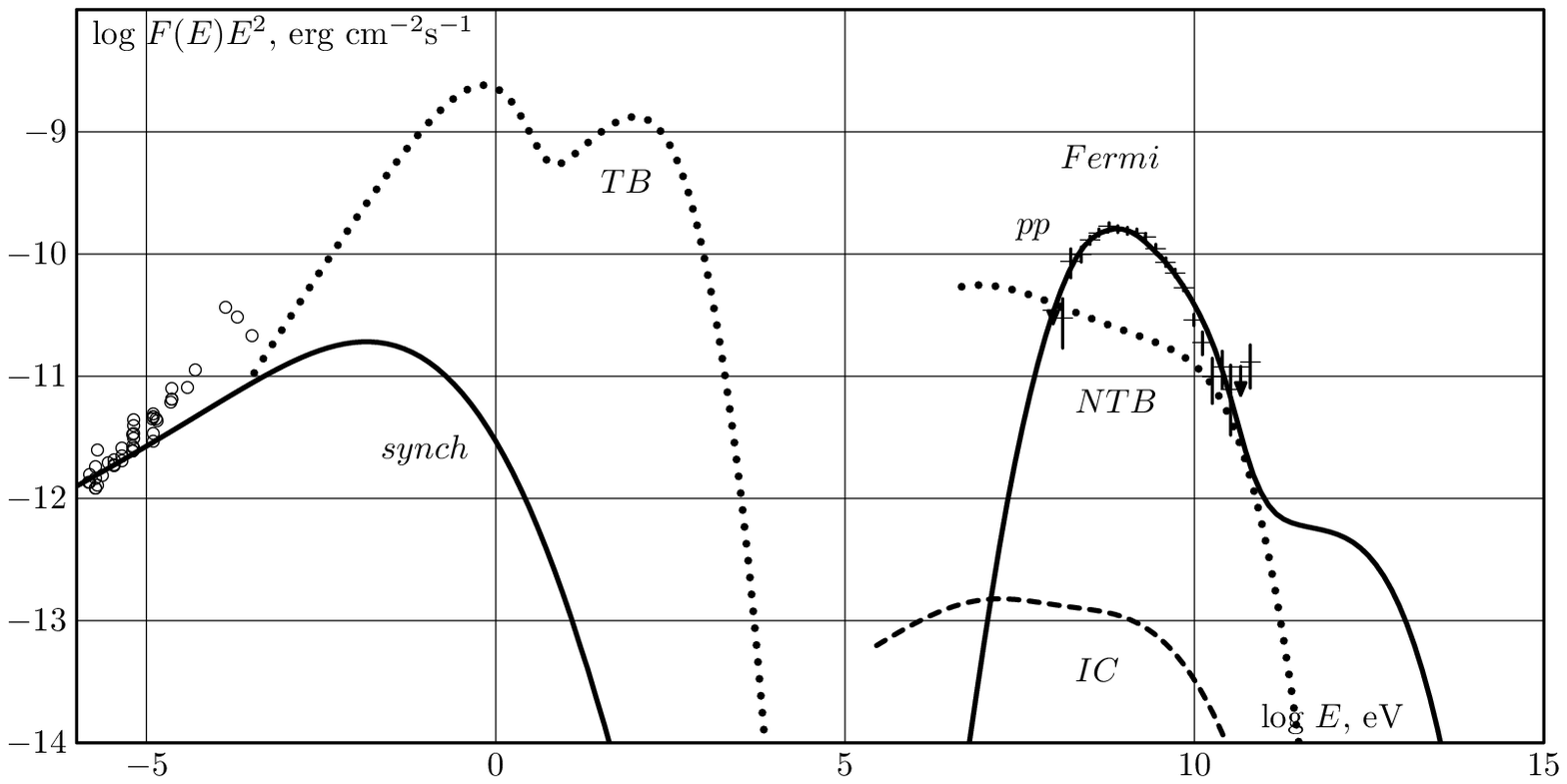}
\caption{The results of modeling of electromagnetic radiation of
W44. The following radiation processes are taken into account:
synchrotron radiation of accelerated electrons (solid curve on the
left), IC emission (dashed line), gamma-ray emission from pion decay
(solid line on the right), thermal bremsstrahlung (dotted line on
the left), nonthermal bremsstrahlung (dotted line on the right).
 Experimental
data in  gamma-rays by the Fermi LAT \citep{ackermann13} (data with
error bars) and in radio bands \citep{castelletti07, arnaud16}
(circles)  are also shown. }
\end{figure}

\begin{figure}
\includegraphics[width=8.0cm]{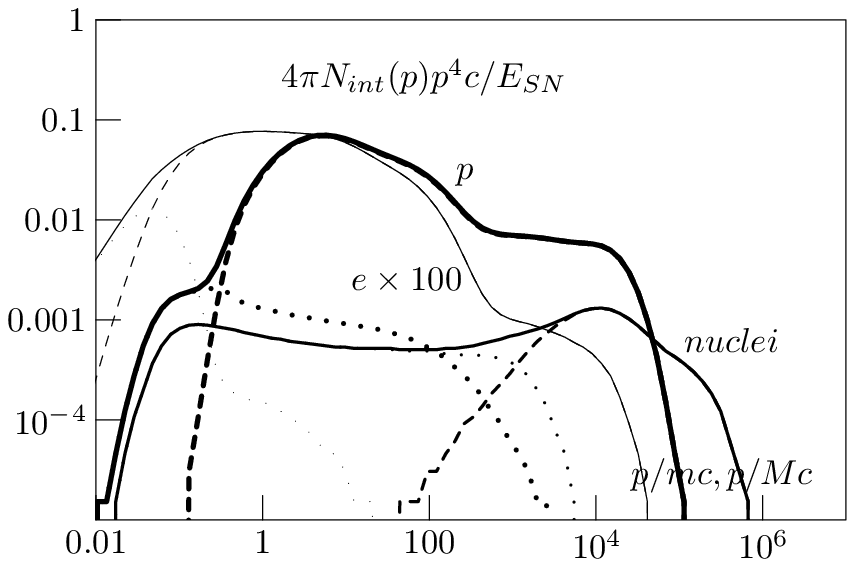}
\caption{ Spectra of protons (thick lines) and electrons (thin
lines) produced at the forward shock,
 and nuclei (normal lines)  produced at the reverse shock in SNR of Type IIP
during $400$ kyr after explosion. The spatially integrated spectra
of particles (dotted lines), the spectra of particles escaped from
the remnant (dashed lines), and the sum (solid lines) are shown.}
\end{figure}



Results of multi-band modeling of SNR W44 are shown in Fig. 3.
Thermal emission has two components. One is produced by the hot gas
in the remnant interior while the lower energy component is produced
by the dense gas that cooled and recombined behind the shock front.
This gas
 produces a significant amount of the thermal radio emission that dominates the synchrotron
 radio emission at high frequencies.

The spectra of particles $N_{int}$ produced during  400 kyr after
supernova explosion
 are shown in Fig.4. They are calculated via the integration throughout the simulation domain and via the
integration on time of  the outward diffusive flux at the simulation
boundary at $r=2R_f$. About $26\%$ of the kinetic energy of the
explosion is transferred to cosmic rays. Almost all this energy is
gone with escaped particles. Note that almost all protons and
electrons accelerated at the forward shock have left the remnant.
This is because the neutral damping of MHD waves, which confine
cosmic rays, was taken into account in the downstream region. In
this regard, the proton and electron  spectra shown in Fig.4 are the
source
 spectra of galactic CRs. This is not so for the ion spectrum. A significant part of
ions accelerated at the reverse shock are still confined in the
central part of the remnant, where the gas is fully ionized. The
exit of particles from the ionized regions of SNR is regulated by
another kind of damping of MHD waves that is not considered here.

As was mentioned before the acceleration efficiency increases just
before and after the transition to the radiative stage. The
corresponding cosmic ray spectra of protons and electrons shows a
spectral hardening at several
 hundreds GeV. The effect is stronger for electrons because their injection rate increases with time according
 to Eq. (13).

\begin{figure}
\begin{center}
\includegraphics[width=8.0cm]{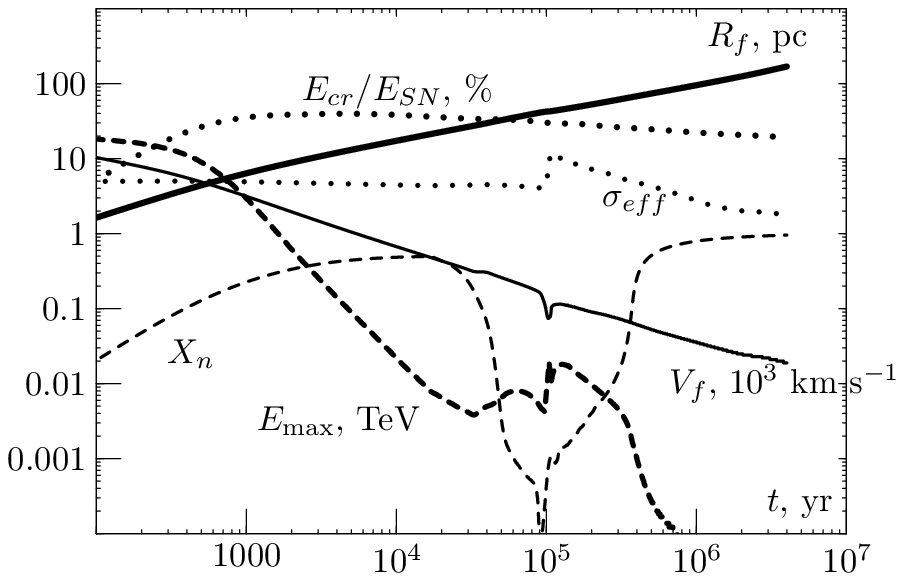}
\end{center}
\caption{ Dependence on time of the forward shock radius $R_f$
(thick solid line), the shock speed $V_f$
 (thin solid line), the effective shock compression ratio $\sigma _{eff}$ (thin dotted line),
the maximum energy of particles accelerated at forward shock
$E_{\max }$ (dashed line), the fraction of explosion energy
transformed into cosmic rays $E_{cr}/E_{SN}$ (thick dotted line) and
neutral fraction $X_n$ (thin dashed line) calculated for SNR of Ia
supernovae.}
\end{figure}

\section{Modeling of DSA in SNR of Type Ia supernova}

The parameters of supernova modeling are given in Table I. The
explosion parameters were adjusted to reproduce
 multiwave observations of young SNR Tycho. The distance to this SNR is very uncertain. So we fix the explosion energy
 to a  value $E_{SN}=1.2\cdot 10^{51}$ erg of the one-dimensional delayed detonation
model of the Tycho supernova explosion \citep{badenes06}. Then the
distance and the ambient number density $n_H$ were adjusted to
reproduce the age of SNR and its angular diameter of $8'$. The
parameter $h_w=0.7$ in Eqs. (3),(5) and (8) was adjusted to
reproduce the observable gamma-ray spectrum.

It was assumed that about 5 solar masses of Hydrogen is ionized
before and during the supernova explosion.

The remnant evolves in a low-density medium. That is why the
transition to the radiative stage occurs at $100$ kyrs. The regular
magnetic field produces a stronger limitation of $\sigma
_\mathrm{eff}\sim 10$ at this stage.

\begin{figure}
\includegraphics[width=8.0cm]{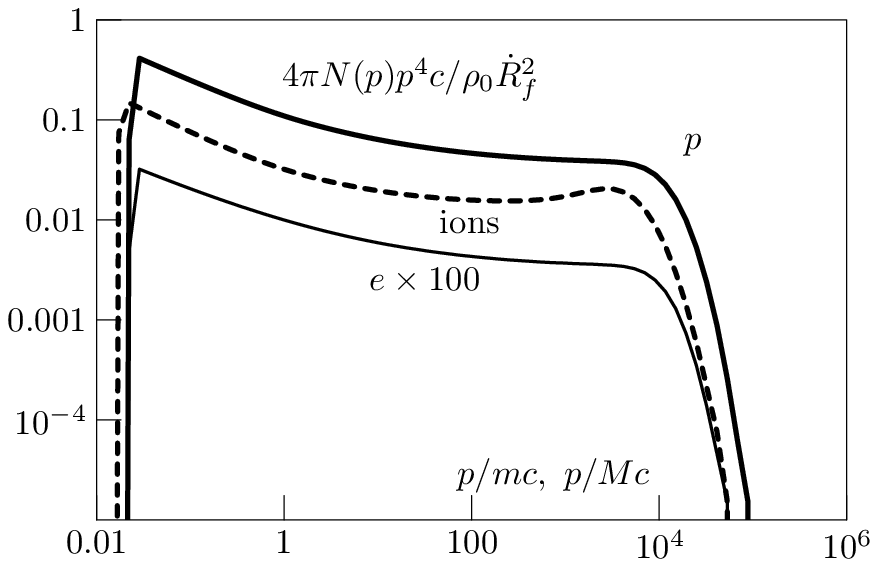}
\caption{The spectra of particles in Tycho SNR. Proton and electron
spectra at the forward shock and spectrum of ions accelerated at the reverse
shock are shown. }
\end{figure}

Spectra of accelerated in Tycho SNR protons, ions, and electrons at
present
 are shown in Fig.6. They are rather soft due to  the Alfv\'enic transport
in the downstream region. This also provides a good agreement with
radio, X-ray,
 and gamma-ray data (see Fig.7).

The spectra of particles $N_{int}$ produced during  4 Myr after
supernova explosion
 are shown in Fig.8. They are calculated via the integration throughout the simulation domain and via the
integration on time of  the outward diffusive flux at the simulation
boundary at $r=2R_f$. About $19\%$ of the kinetic energy of the
explosion is transferred to cosmic rays. Almost all this energy is
gone with escaped particles. The maximum energy of escaped particles
is 50 TeV for this SNR. Similar to the case of IIP type SNR almost
all
 accelerated protons and electrons left the remnant and their spectra can be considered as source spectra
of galactic cosmic rays.

The energy of the hardening is lower for Ia Type SNRs. Note that
transition to the radiative stage occurs when the shock radius
 is 40 pc. Probably in many cases, the shock will collide with a denser medium before the transition.
Then the situation will be similar to the one considered in the
previous Section.

\begin{figure}
\includegraphics[width=8.0cm]{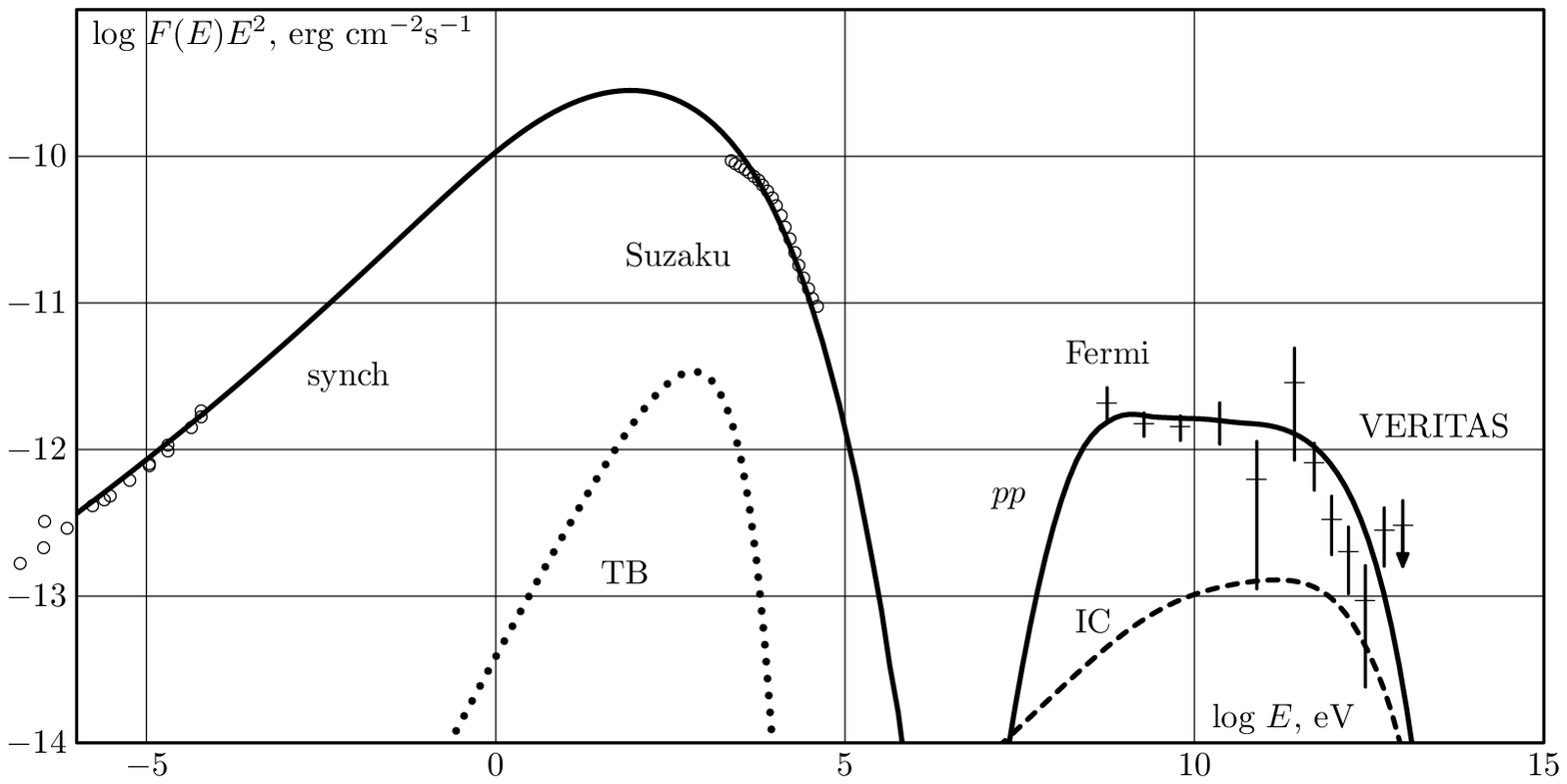}
\caption{The results of modeling of electromagnetic radiation of
Tycho SNR. The following radiation processes are taken into account:
synchrotron radiation of accelerated electrons (solid curve on the
left), IC emission (dashed line), gamma-ray emission from pion decay
(solid line on the right), thermal bremsstrahlung (dotted line on
the left).
 Experimental
data in  gamma-rays by the Fermi LAT and VERITAS
\citep{archambault17};  (data with error bars), radio
\citep{klein79}, and analytical approximation of continuum X-rays by
Suzaku \citep{tamagawa09} (circles)  are also shown. }
\end{figure}

\begin{figure}
\includegraphics[width=8.0cm]{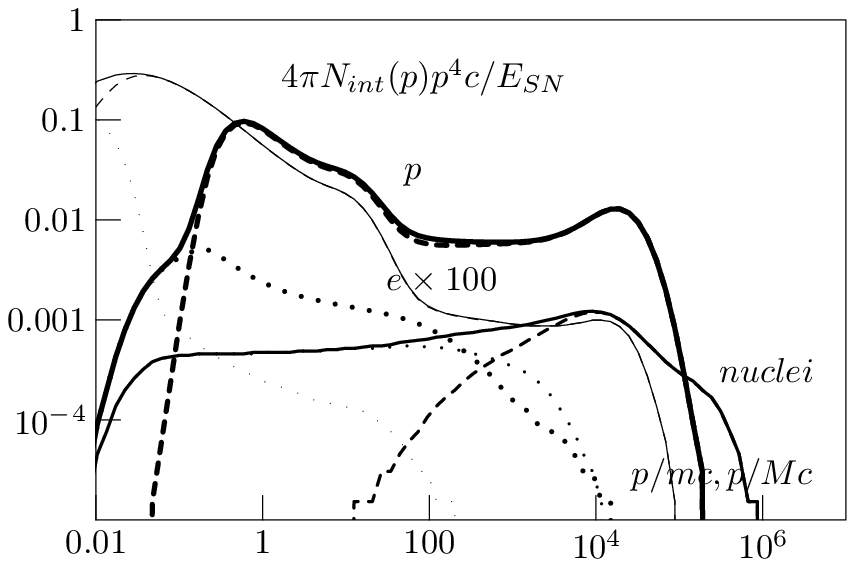}
\caption{ Spectra of protons (thick lines) and electrons (thin
lines) produced at the forward shock,
 and nuclei (normal lines)  produced at the reverse shock in SNR of Type Ia
during $4$ Myr after explosion. The spatially integrated spectra of
particles (dotted lines), the spectra of particles escaped from the
remnant (dashed lines), and the sum (solid lines) are shown.}
\end{figure}

\section{Discussion}

The transition to the radiative stage begins when the radiative
losses
 are comparable with adiabatic losses in the downstream region (see the corresponding terms in Eq. (3)).
Assuming a linear profile of the gas velocity we get

\begin{equation}
\Lambda (T_e)n_H^2\sigma ^2\sim \frac {(\sigma -1)\gamma _g\xi
_g\rho V_f^3}{(\gamma _g-1)\sigma R}.
\end{equation}
Here $\xi _g=P_g/\rho V_f^2$ is the ratio of the gas pressure just
downstream of the shock $P_g$ to the shock ram pressure $\rho
V_f^2$, $\sigma $ is the shock compression ratio. Using the relation

\begin{equation}
\rho V_f^2R_f^3=0.2E_{SN}
\end{equation}

at the Sedov stage and numeric values $\sigma =4$, $\gamma _g=5/3$,
and $\xi _g=0.75$
 we can obtain the shock speed $V_{\mathrm{rad}}$ at the time of transition

\begin{equation}
V_{\mathrm{rad}}=250\ \mathrm{km\ s}^{-1}n_H^{2/11} \left( \frac {E
_{SN}}{10^{51}\mathrm{erg}}\right) ^{1/11} \left( \frac \Lambda
{10^{-22}\mathrm{erg\ cm}^{3}\mathrm{s}^{-1}}\right) ^{3/11}
\end{equation}

In the absence of damping the maximum energy $E_{\max }$ of
accelerated particles can be found from the condition that the
magnetic field amplified by the cosmic ray streaming instability has
enough
  time to grow from the initial value of $B_b$, that is $\Gamma _{cr}T=\ln (B/B_b) \sim 10$ during the age $T$.

In young SNRs the streaming instability is non-resonant while
in older remnants it is resonant. In spite of the different nature
 the rates of the resonant and non-resonant instabilities are given by similar expressions. For the estimate we use
the non-resonant instability rate $\Gamma _{cr}$ \citep{bell04}

\begin{equation}
\Gamma _{cr}= \frac {J_{\mathrm{el}}B}{2c\rho v_A}, \
\end{equation}
where $J_{\mathrm{el}}$ is the electric current of the highest energy cosmic
rays escaping into the upstream region. This results in expression
\citep{zirakashvili08}

\begin{equation}
E_{\max}=\frac {\eta _{\mathrm{esc}}qBV_f^3T}{4cv_A\ln (B/B_b) }= 2\
\mathrm{TeV}\left( \frac {\eta _{\mathrm{esc}}}{0.05}\right) \left( \frac
{V_f}{10^3\mathrm{km\ s}^{-1}}\right) ^2
m_{\mathrm{exp}}n_H^{0.5}R_{\mathrm{pc}}.
\end{equation}
Here $\eta _{esc}$ is the ratio of the energy flux of highest energy
particles to the shock energy flux $\rho V_f^3/2$ and
 $m_{\mathrm{exp}}=V_fT/R_f$ is the expansion parameter of the shock.
At the beginning of the Sedov stage $V_f^2=2E_{SN}/M_{ej}$,
$R_f=(3M_{ej}/4\pi m\cdot 1.4n_H)^{1/3}$ and
\[
E_{\max}= 2\left( \frac{3}{4\pi \cdot 1.4}\right) ^{1/3} \frac
{m_{\mathrm{exp}}\eta
_{\mathrm{esc}}qBE_{SN}}{4m^{1/3}n_H^{1/3}M^{2/3}_{ej}cv_A\ln (B/B_b) }=
\]
\begin{equation}
400\ \mathrm{TeV}\left( \frac {\eta _{\mathrm{esc}}}{0.05}\right) \left(
\frac {E_{SN}}{10^{51}\ \mathrm{erg}}\right) \left( \frac
{M_{ej}}{M_{\odot}}\right) ^{-2/3} m_{\mathrm{exp}}n_H^{1/6}.
\end{equation}
The value $\eta _{\mathrm{esc}}=0.05$ corresponds to the effectively
accelerating shock with $E^{-2}$ spectrum of particles and CR
pressure of the order $50\%$ of the shock ram pressure. The
Alfv\'enic transport downstream results in steeper spectra, lower CR
pressure (see Fig.6), and lower $\eta _{\mathrm{esc}}\sim 0.01$. Then for
parameters of IIP and Ia Type supernovae (see Table I) and
$m_{\mathrm{exp}}=0.5$ we get the maximum energies of 20 TeV and 30
TeV at the beginning of the Sedov stage
 in qualitative agreement with Fig.1 and Fig.5.

So the maximum energies determined by CR streaming instability are
not higher than 100 TeV in SNRs considered. Higher energies can be
reached
  for SNRs shocks propagating in rarefied bubbles created by Type Ib/c
supernova progenitors where the medium is prepared
 for the efficient DSA \citep{zirakashvili18b,zirakashvili21}
or in dense progenitor winds of IIn Type  SNRs
\citep{zirakashvili16b}.

The maximum energy at the instant of transition to the radiative
stage $E_{\mathrm{rad}}$ can be found from equations (16),(17), (19)
\begin{equation}
E_{\mathrm{rad}}=200\ \mathrm{GeV}\left( \frac {\eta _{\mathrm{esc}}}{0.01}\right)
n_H^{9/22} \left( \frac {E _{SN}}{10^{51}\mathrm{erg}}\right)
^{5/11}
\left( \frac \Lambda
{10^{-22}\mathrm{erg\ cm}^{3}\mathrm{s}^{-1}}\right) ^{4/11}
\end{equation}
 where the value $m_{\mathrm{exp}}=0.4$ was used.
This gives the energy of hardening 100 and 600 GeV for Ia and IIP
Type SNRs.

It is expected that the effect will be similar for light cosmic ray
nuclei. However
 details of the ionization, e.g. a high ionization potential of the Helium can result in
  some peculiarities of the hardening.

We expect that the effect of the hardening can be different for heavy
nuclei. The matter is that the nuclei injected
 into DSA are single or double charged. The further ionization occurs via collisions with thermal particles
(a so called stripping) and a photoionization. The photoionization
is possible when accelerated particles reach high
 Lorentz factors and interact with optical, infrared and microwave background photons \citep{morlino11}.
For example, it takes $\sim 10^5$ years for a full ionization of
Iron nuclei with a Lorentz factor 100 accelerated in SNR evolving in
the dense gas with a number density of $10\ \mathrm{cm}^{-3}$ (see
Fig.1 of \citep{morlino11}). This time is higher than the age of the
transition to the radiative stage $\sim 10^4$ years. Actually the
heavy nuclei have the time for the ionization up to the charge state
$\sim 10$. They will be stripped further after the end of the
acceleration and during a propagation in the Galaxy. Therefore it is
expected that heavy nuclei have lower observable rigidities of the
hardening.

It is important to note that high-energy measurements of protons and
nuclei energy spectra in the cosmic ray experiment AMS confirm the
earlier experimental results of ATIC-2, CREAM and PAMELA
measurements on the presence of spectral hardening at magnetic
rigidity at about 200 GV
 (see e.g.review of \citet{serpico18} and references therein). Some peculiarities
for the hardening of Iron nuclei
 were also reported \citep{schroer21}.

The nature of this hardening is not clear yet. In principle it may
reflect the source spectra or the peculiarity of the energy
dependence of cosmic ray leakage time from the Galaxy.

In our modeling the hardening at energies $100-500$ GeV in the
source proton and electron spectra
 is related with a higher acceleration efficiency after the full reonization of the medium just before
the beginning of the radiative stage and with the strong gas
compression behind the shock at the radiation stage. Probably this
spectral feature is also presented in the energy dependence of the
cosmic ray leakage time because of the self-confinement of cosmic
rays (see a review of \citet{blasi19} and references therein). In
self-confinement models, cosmic rays generate MHD waves via cosmic
ray streaming instability. These waves in turn scatter cosmic ray
particles and regulate their diffusion and confinement in the
Galaxy. The streaming instability is produced mainly by protons and
$\alpha $-particles. So the hardening in their source spectra  will
result in a change of the energy dependence of the interstellar
diffusion coefficient.

\section{Conclusion}
Our results and conclusions are the following:

1) We performed the modeling of particle acceleration in SNRs up to
the late stages of the remnant
 evolution when almost all particles accelerated at the forward shock have left the remnant.

2) We show that transition to the radiative phase of supernova
remnants
  is accompanied by a higher acceleration
 efficiency of cosmic rays. It leads to the  hardening of the cosmic ray proton and electron source spectrum at energies
of several hundreds GeV. This energy is simply the maximum energy of
accelerated particles at the time of the transition. Cosmic ray
particles at lower energies are mainly accelerated
 and escaped the remnant at the radiative stage.

3) The effect is stronger for cosmic ray electrons because of the
higher electron injection rate in old SNRs.

4) We expect that the effect is different for heavy nuclei because
of the partial ionization.

\section*{Acknowledgements}
The work was partially supported by the Russian Foundation for Basic
Research grant 19-02-00043. The work was also partly performed at the Unique scientific
installation "Astrophysical Complex of MSU-ISU"
(agreement 13.UNU.21.0007). We also thank the referee Luke
Drury for valuable comments. 

\section*{Data Availability}

All results in this paper were obtained using available published
data.









\bsp    
\label{lastpage}
\end{document}